\documentclass[aps,prl,twocolumn,groupeaddress,showpacs]{revtex4-1}

\usepackage{graphicx}
\usepackage{bm}
\usepackage{color}
\begin{document}

\preprint{APS/123-QED}

\title[Sample title]{Direct Observation of Superconductivity in Calcium-Intercalated
Bilayer Graphene by {\it in situ} Electrical Transport Measurements
}

\author{S. Ichinokura$^{1,*}$, K. Sugawara$^{2}$, A. Takayama$^{1}$, T. Takahashi$^{2,3}$, and S. Hasegawa$^{1}$}

\affiliation{$^1$Department of Physics, University of Tokyo, 7-3-1 Hongo, Bunkyo-ku, Tokyo 113-0033, Japan} 
\affiliation{$^2$WPI Research Center, Advanced Institute for Materials Research, Tohoku University, Sendai 980-8577, Japan} 
\affiliation{$^3$Department of Physics, Tohoku University, Sendai 980-8578, Japan}

\date{\today}

\begin{abstract}
We report the superconductivity in Ca-intercalated bilayer graphene C$_6$CaC$_6$, 
the thinnest limit of Ca graphite intercalation compound. 
We performed \textit{in situ} electrical transport measurements on pristine bilayer graphene, 
C$_6$LiC$_6$ and C$_6$CaC$_6$ fabricated on SiC substrate under zero and non-zero magnetic field. 
While both bilayer graphene and C$_6$LiC$_6$ show non-superconducting behavior, 
C$_6$CaC$_6$ exhibits the superconductivity with transition temperature ($T_{{\rm c}}$) of 4.0 K.  
The observed $T_{{\rm c}}$ in C$_6$CaC$_6$ and the absence of superconductivity in C$_6$LiC$_6$ show a good agreement 
with the theoretical prediction, 
suggesting the importance of a free-electron-like metallic band at the Fermi level to drive the superconductivity.
\end{abstract}

\maketitle
Atomic-layer superconductors (ALSCs), where only one or a few atomic layers at the surfaces or interfaces become superconducting\cite{Qin, Zhang2, Uchihashi1, Yamada, Brun, Reyren1, ZHANG, Ge}, 
have attracted considerable attentions owing to their two-dimensionallity\cite{Brun, Reyren1, ZHANG}
and possible existence of symmetry mixed Cooper pairs\cite{Gorkov}.
It is known that in ALSCs fabricated on semiconductor substrates,
the superconducting-transition temperature ($T_{{\rm c}}$) is in general suppressed in comparison
with that of bulk\cite{Qin, Zhang2, Uchihashi1, Yamada, Brun}
due to the interference from the substrate except for limited cases\cite{ZHANG, Ge}. 
In this sense, self-standing ALSCs free from the substrate effect have been desired to pursue higher $T_{{\rm c}}$.

Graphene is a single atomic sheet of graphite and now a target of intensive theoretical and experimental studies 
because of its various peculiar but attractive properties such as the massless nature of carriers\cite{Novoselov1}, 
the anomalous quantum Hall effect\cite{Novoselov1}, 
the high carrier mobility to drive the ballistic transport\cite{Novoselov3}, {\it etc.} 
Besides these superior properties, graphene has a very important inherent property, 
namely, self-standing crystal structure basically isolated from substrate.
In this context, graphene is regarded as one of ideal materials to realize ALSCs free from the substrate effect. 
Just after the discovery of graphene\cite{Novoselov4}, 
many intensive efforts have been made to fabricate superconducting graphene 
by intercalating metals like in graphite intercalation compounds (GICs).
Ca-intercalated bilayer graphene (C$_6$CaC$_6$) is regarded as the most promising candidate 
because corresponding Ca-intercalated GIC (C$_6$Ca) has the highest $T_{{\rm c}}$ of 11.5 K among all GICs\cite{Weller,Emery,jobiliong2007anisotropic, Xie200643}.
While some theoretical\cite{Mazin, Jishi} and spectroscopic\cite{Kanetani} studies have proposed 
a similar electronic structure between C$_6$Ca\cite{Csanyi, Sugawara, Yang} and C$_6$CaC$_6$, 
suggestive of superconductivity in C$_6$CaC$_6$, no direct evidences for superconductivity, 
such as superconducting gap and/or zero-resistance, have been reported. 
This may be mainly due to difficulty in handling reactive samples containing Ca, 
requiring an \textit{in situ} measurement under ultrahigh vacuum at ultralow temperature\cite{Zhang2, Uchihashi1, Yamada, Brun}.

In this Letter, we report on the observation 
of the zero-resistance state in C$_6$CaC$_6$ to directry prove the occurrence of supercondutivity 
by \textit{in situ} electrical transport measurements.
We compare the experimental results with those for pristine bilayer graphene and C$_6$LiC$_6$
as well as the theoretical predictions to obtain an insight into the superconducting property 
and mechanism in intercalated bilayer graphene.

\begin{figure}[t] 
\includegraphics[width=3.4in]{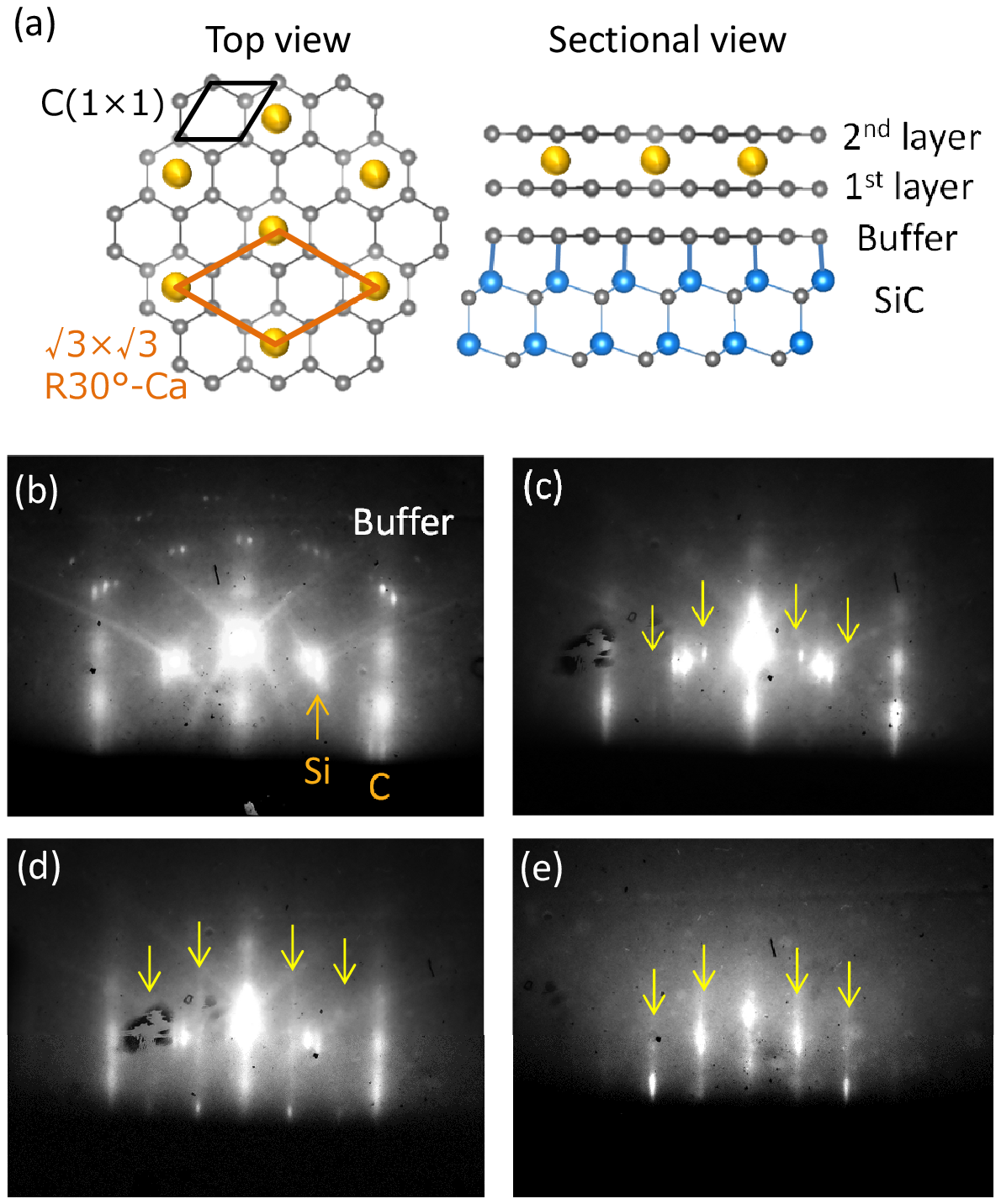}　
\caption{(Color Online) (a) Crystal structure of C$_6$CaC$_6$ on SiC substrate.
(b) RHEED pattern of pristine bilayer graphene. Si(1$\times $1) spots from the substrate 
and C(1$\times $1) spots from graphene are shown. 
Spots from the buffer layer with 6$\sqrt3 \times 6\sqrt3$ R30{$^\circ$} periodicity are also seen.
(c) RHEED pattern of C$_6$LiC$_6$. (d), (e) Same for C$_6$CaC$_6$ 
(d) after the first Li-Ca replacing treatment and 
(e) after several cycles of the replacing treatments. 
Yellow allows indicate $\sqrt3 \times \sqrt3$ R30{$^\circ$} spots and streaks.
}　
\label{fig1}　
\end{figure}%
\begin{figure}[t] 
\centering 
\includegraphics[width=5cm]{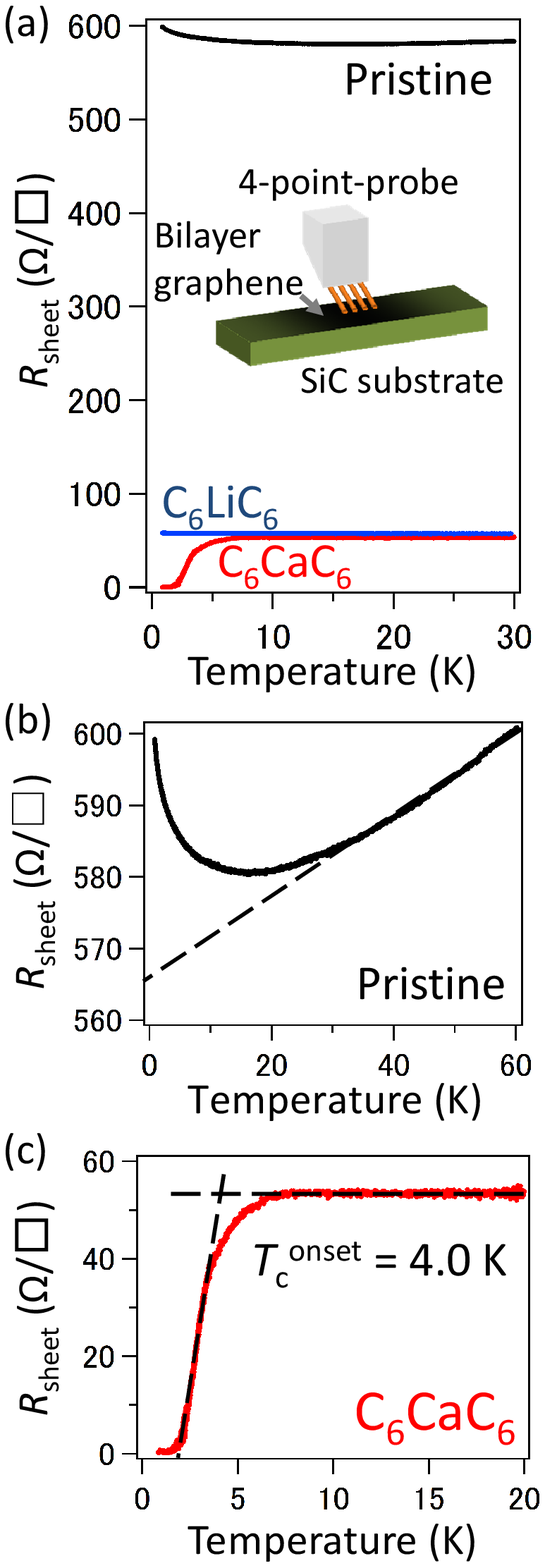}　
\caption{(Color Online) (a) Sheet resistance $R_{{\rm sheet}}$ of pristine bilayer graphene, 
C$_6$LiC$_6$ and C$_6$CaC$_6$ as a function of temperature.
Inset shows a schematic picture of 4-point-probe measurement set-up.
(b) $R_{{\rm sheet}}$ of pristine bilayer graphene in an expanded scale from 0.8 to 60 K.
Dashed line shows extrapolation toward 0 K. 
(c) $R_{{\rm sheet}}$ of C$_6$CaC$_6$ from 0.8 K to 20 K,
 showing $T_{{\rm c}}^{{\rm onset}}$ = 4.0 K and $T_{{\rm c}}^{{\rm zero}}$ = 2.0 K.
}　
\label{fig2}　
\end{figure}%
\begin{figure}[t] 
\includegraphics[width=3.4in]{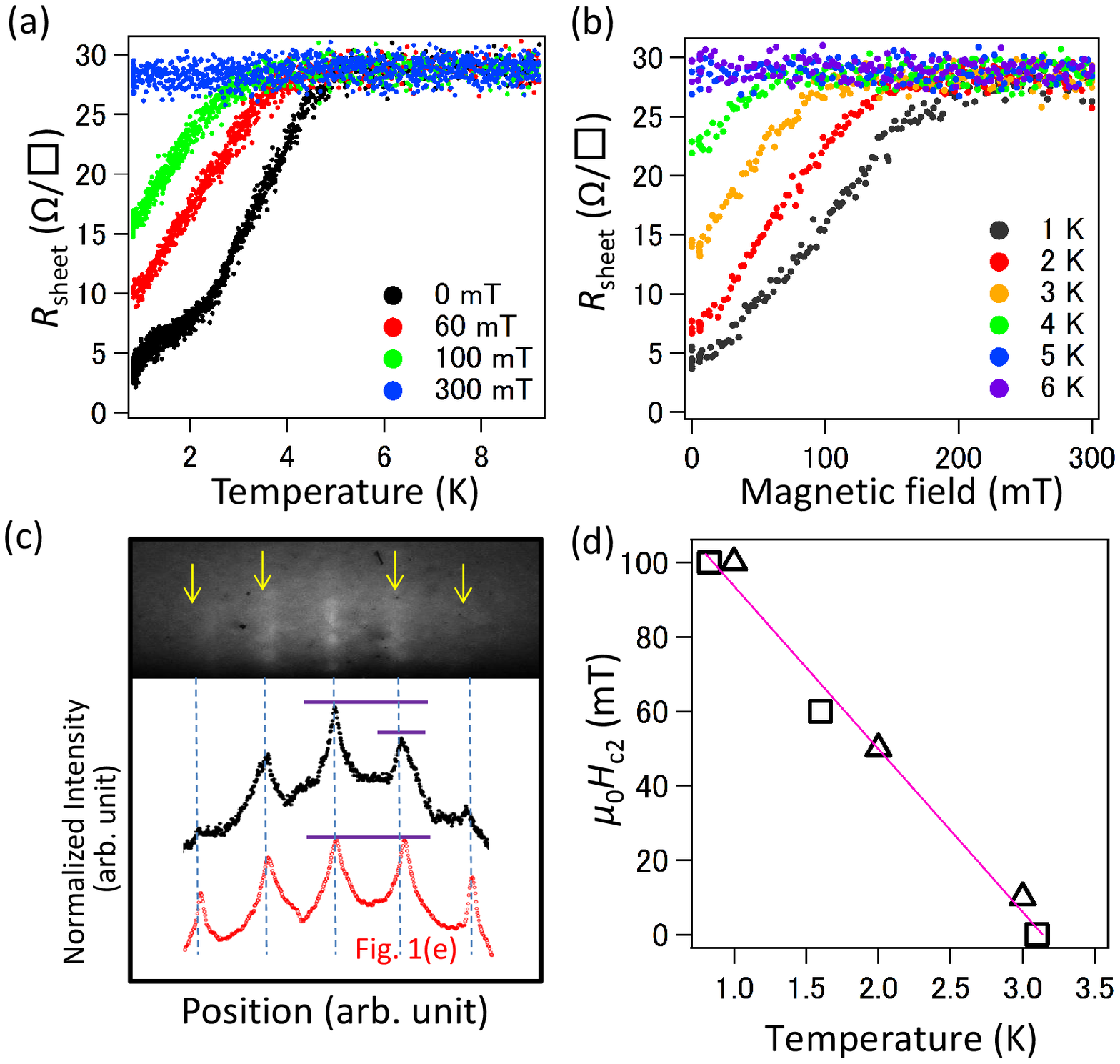}　
\caption{(Color Online) (a) $R_{{\rm sheet}}$ of C$_6$CaC$_6$ as a function of temperature 
under different magnetic fields. 
(b) Same as (a) as a function of magnetic field for different temperatures. 
Magnetic field was applied perpendicular to the sample surface. 
(c) (Picture) RHEED pattern of this sample, 
showing weaker $\sqrt3 \times \sqrt3$ R30{$^\circ$} streaks than those in Fig. 1(e).  
Note that the sample is different from that in Fig. 1(e). 
(Plot) The intensity of RHEED pattern as a function of horizontal position 
for Fig. 3(c) (upper black dot) and Fig. 1(e) (lower red circle). 
The intensity is normalized by the peak height of central ($1 \times 1$) streak. 
(d) Temperature dependence of the upper critical field $\mu _{0}${\it H}$_{c2}$ 
obtained from (a) (squares) and (b) (triangles). 
$\mu _{0}${\it H}$_{c2}$ is defined as the magnetic field 
where $R_{{\rm sheet}}$ drops to a half of normal-state resistance. 
Solid line shows the fitting based with the Ginzburg-Landau theory[40].
}　
\label{fig3}　
\end{figure}%

To fabricate intercalated bilayer graphene [Fig. 1(a)], 
we at first prepared a high-quality
pristine bilayer graphene sheet on a {\it n}-type Si rich 6{\it H}-SiC(0001) single crystal 
by heating the crystal up to 1550{$^\circ$C} in an argon atmosphere\cite{Sugawara2}.
After a short exposure to air, the grown bilayer graphene/SiC was transferred  
into a RHEED (reflection-high-energy electron diffraction) vacuum chamber 
where the transport measurements were also done. 
After heating the sample at 400{$^\circ$C} for several hours in the chamber,
we observed a RHEED pattern typical of clean bilayer graphene on SiC[Fig. 1(b)].
Then, we deposited Li atoms on the bilayer graphene sheet using a SAES Getter dispenser. 
The Li deposition produced several sharp $\sqrt3 \times \sqrt3$ R30{$^\circ$} spots 
in the RHEED pattern [pointed by yellow arrows in Fig.1(c)], 
indicating that Li atoms are intercalated in a regular manner 
between two adjacent graphene layers\cite{Sugawara3} as in bulk C$_6$Li. 
After confirming the growth of high-quality C$_6$LiC$_6$ on SiC, 
we then deposited Ca atoms on this C$_6$LiC$_6$ sheet to replace Li with Ca.  
During the Ca deposition, we kept the substrate at 150{$^\circ$C},
slightly above the Li desorption temperature of 145{$^\circ$C}. 
We observed that the Ca deposition transformed the $\sqrt3 \times \sqrt3$ R30{$^\circ$} spots 
into streaks [Fig. 1(d)],
suggesting that intercalated Li atoms are replaced by Ca atoms\cite{Kanetani}.
Repeated cycles of Li and Ca deposition together with annealing at appropriate temperatures
made the $\sqrt3 \times \sqrt3$ R30{$^\circ$} streaks brighter and brighter. 
Finally, we obtained the RHEED pattern in Fig. 1(e),
where we observe sharp $\sqrt3 \times \sqrt3$ R30{$^\circ$} streaks,
indicative of well-ordered C$_6$CaC$_6$.

After fabricating the sample as described above, 
we then transferred it on a stage for \textit{in situ} electrical transport measurements 
with the 4-point-probe (4PP) method in the same vacuum chamber\cite{yamada2012surface}.
As shown in Fig. 2(a), 
the 4PP chip composed of four copper wires in 100 $\mu $m $\phi $
was contacted on the sample, and then cooled down to 0.8 K together with the sample. 
The results of temperature-dependent transport measurements 
on bilayer graphene, C$_6$LiC$_6$ and C$_6$CaC$_6$
are compared in Fig. 2(a).
The sheet resistance ($R_{{\rm sheet}}$) of pristine bilayer graphene 
shows a metallic character above 20 K, 
but turns into an insulating one at lower temperatures. 
This is usually observed in the transport characteristics of graphene 
and interpreted in terms of the localization effect due to the quantum interference\cite{Wu,Tikhonenko1,Chen}.
The residual resistance at 0 K is estimated to be 565 $\Omega $ [Fig. 2(b)].
In C$_6$LiC$_6$ and C$_6$CaC$_6$, on the other hand, 
the $R_{{\rm sheet}}$ behaves as metallic, 
and their resistance is as low as ca. 10\% of that of pristine bilayer graphene.
This is because of the increase of the Fermi-surface volume 
by the career doping from Li or Ca atoms 
and the resultant folding of Brillouin zone 
due to the induced $\sqrt3 \times \sqrt3$ R30{$^\circ$} superstructure\cite{Kanetani}.
Importantly, the transition of $R_{{\rm sheet}}$ to zero-resistance 
is clearly seen at around 2 K in C$_6$CaC$_6$, 
which is the direct evidence of occurrence of superconductivity.
It is also noteworthy that C$_6$LiC$_6$ does not show superconductivity down to 0.8 K, 
and instead exhibits a weak localization behavior as evident from a slight upturn in resistance 
at low temperature.
Figure 2(c) shows the resistance of C$_6$CaC$_6$ at temperatures from 0.8 K to 20 K, 
highlighting that $R_{{\rm sheet}}$ suddenly drops at around 4 K 
($T_{{\rm c}}^{{\rm onset}}$ = 4.0 K). One can also notice in Fig. 2(c) that $R_{{\rm sheet}}$ 
starts to decrease even above 4 K. 
This is probably due to the superconducting fluctuation 
inherent to low-dimensional superconductors\cite{Aslamasov1968238}.

To obtain further evidence for superconductivity,
we conducted magnetoresistance measurements on another C$_6$CaC$_6$ sample
prepared with a same method as described above.
In the experiments, the magnetic field was applied perpendicular to the surface.
Figures 3(a) and (b) show $R_{{\rm sheet}}$ data as a function of temperatures and magnetic fields.
As seen in Fig. 3(a), the $T_{{\rm c}}^{{\rm onset}}$ gradually shifts to lower temperature 
as the magnetic field is increased.
A similar behavior of $T_{{\rm c}}^{{\rm onset}}$ is also seen in Fig. 3(b).
It is noted that the $R_{{\rm sheet}}$ curve did not fully drop to zero-resistance even at 0.8 K 
under zero magnetic field, suggesting that only a limited part of sample became superconducting 
and the superconducting paths were not fully connected between the current probes in this sample.
In fact, as shown in Fig. 3(c), the $\sqrt3 \times \sqrt3$ R30{$^\circ$} streak 
originating in C$_6$CaC$_6$ is weaker in the present sample than in that of Fig. 1(e).  
This in return supports that the superconductivity emerges 
in the well-ordered $\sqrt3 \times \sqrt3$ R30{$^\circ$} superstructure.
To see the temperature-dependence of the upper critical field ($\mu _{0}${\it H}$_{c2}$), 
we plot the magnitude of magnetic field 
at which the $R_{{\rm sheet}}$ is a half of the normal-state-resistance 
as a function of temperature in Fig. 3(d). 
We find that extracted points are well aligned linearly as seen in Fig. 3(d), 
indicating that these experimental results can be analyzed 
in the framework of the Ginzburg-Landau (GL) theory\cite{tinkham2012introduction}.
Numerical fittings with the GL theory show 
that the in-plane GL coherence length at zero Kelvin $\xi $(0) is 49 $\pm $ 1 nm.
This is comparable to or slightly larger than the in-plane  $\xi $(0) reported for bulk C$_6$Ca 
(29 -- 36 nm)\cite{Weller,Emery,jobiliong2007anisotropic, Xie200643}.
This shows a striking contrast with the case of usual ALSCs such as In on Si(111), 
where the $\xi $(0) $\sim $ 25 nm\cite{Yamada} is much shorter than that of bulk (250 -- 440 nm)\cite{PhysRev.139.A1482}. 
This suggests the strong two-dimensional nature of superconductivity in both C$_6$CaC$_6$ and C$_6$Ca. 

Now we discuss the origin of superconductivity in C$_6$CaC$_6$. 
While the $T_{{\rm c}}$ is lower than that of bulk C$_6$Ca (11.5 K), 
the present observation of superconductivity in C$_6$CaC$_6$ is striking. 
Because, no superconductivity has been observed so far in second-stage GICs (C$_{12}$A 
where A is an alkali metal) which are a bulk analog of intercalated bilayer graphene (C$_6$AC$_6$).
A simple estimation indicates that the carrier density in a single carbon layer in C$_6$CaC$_6$
should be reduced to a half of that in bulk C$_6$Ca. 
However, Mazin and Balatsky\cite{Mazin} have theoretically proposed 
the possibility of superconductivity in C$_6$CaC$_6$ despite the reduced carrier density, 
based on the similarity in the topology of Fermi surfaces 
consisting of a nearly-free electron band at the center of Brillouin zone and the $\pi $* band 
at the zone boundary. They showed by calculations 
that the nearly-free electron band is partially occupied in C$_6$CaC$_6$ 
while it is totally unoccupied in C$_6$LiC$_6$, 
indicating the importance of the metallic nearly-free electron band for the superconductivity.  
This theoretical prediction shows a good agreement with the present observation 
that the superconductivity emerges in C$_6$CaC$_6$ while not in C$_6$LiC$_6$.  
Jishi {\it et al.}\cite{Jishi} have confirmed the band structure of intercalated bilayer graphene 
proposed by Mazin and Balatsky and further predicted the $T_{{\rm c}}$ of 4.7 K for C$_6$CaC$_6$.
This shows a good quantitative agreement with the present observation of 
$T_{{\rm c}}$ = 4.0 K. This strongly supports the theoretical proposal\cite{Mazin, Jishi} that the metallic nearly-free electron band at the center of Brillouin zone 
plays an essential role for the superconductivity.  
One may notice here that the calculations\cite{Mazin, Jishi} were done for ideal free-standing 
intercalated bilayer graphene while in reality the intercalated bilayer film 
is fabricated on SiC substrate, which may affect the electronic structure 
and consequently the superconductivity. In fact, a recent scanning tunnel microscope study 
reported that a charge density wave is created at low temperatures in C$_6$CaC$_6$ on SiC 
due to the commensurate lattice matching\cite{Shimizu}.  
It should be noted that despite these interferences from the substrate 
the $T_{{\rm c}}$ observed in C$_6$CaC$_6$ fabricated on SiC shows 
almost the same value as predicted for a free-standing C$_6$CaC$_6$, 
suggesting that the superconductivity in C$_6$CaC$_6$ is very robust.

In conclusion, we performed \textit{in situ} electrical transport measurements 
on Ca-intercalated bilayer graphene C$_6$CaC$_6$ under zero and non-zero magnetic field.  
We have clearly observed that the resistance steeply drops at 4 K and reaches zero at 2 K 
under zero magnetic field, showing the occurrence of superconductivity 
with $T_{{\rm c}}^{{\rm onset}}$ = 4.0 K and $T_{{\rm c}}^{{\rm zero}}$ = 2.0 K. 
The measurement under magnetic field has confirmed 
the superconductivity origin of the observed zero resistance. 
The observed $T_{{\rm c}}^{{\rm onset}}$ is in good agreement with the theoretical prediction. 
The absence of superconductivity in pristine bilayer graphene and Li-intercalated bilayer graphene 
C$_6$LiC$_6$ in contrast to C$_6$CaC$_6$ suggests 
the importance of free-electron-like metallic band at the Fermi level 
to ignite the superconductivity. 

\begin{acknowledgments}
We thank K. Suzuki and T. Sato at Tohoku University for their useful discussion.   
This work was supported by the JSPS (KAKENHI 15H02105, 25246025, 22246006) 
and the MEXT (Grant-in-Aid for Scientific Research on Innovative Areas ``Science of Atomic Layers" 25107003 
and ``Molecular Architectonics" 25110010).
\end{acknowledgments}

$^*$Corresponding author: 

ichinokura@surface.phys.s.u-tokyo.ac.jp

%



\newpage

\end{document}